\documentstyle[12pt,epsf,citesort]{article}
\newcommand{\Z}{{\sf Z \!\!\! Z}}
\newcommand{\R}{{\sf I \!\! R}}
\newcommand{\1}{{\sf 1 \!\! 1}}
\newcommand{\Sign}{\mbox{Sign}}
\newcommand{\p}{\partial}
\newcommand{\e}{\vec e}
\makeatletter
\@addtoreset{equation}{section}
\makeatother

\title{Pions versus Magnons: From QCD to Antiferro- \\
magnets and Quantum Hall Ferromagnets}

\author{O.~B\"ar$^{a}$, M.~Imboden$^b$, and U.-J.~Wiese$^b$
\footnote{on leave from MIT}
\\ \\
$^a$ Institute of Physics, University of Tsukuba \\
Tsukuba, Ibaraki 305-8571, Japan \\
$^b$ Institute for Theoretical Physics, Bern University \\
Sidlerstrasse 5, CH-3012 Bern, Switzerland \\
Preprint UTHEP-479}

\begin{document} 
\maketitle

\vspace{-1cm}

\begin{abstract} \normalsize

The low-energy dynamics of pions and magnons --- the Goldstone bosons of the
strong interactions and of magnetism --- are analogous in many ways. The 
electroweak interactions of pions result from gauging an $SU(2)_L \otimes 
U(1)_Y$ symmetry which then breaks to the $U(1)_{em}$ gauge symmetry of 
electromagnetism. The electromagnetic interactions of magnons arise from 
gauging not only $U(1)_{em}$ but also the $SU(2)_s$ spin rotational symmetry, 
with the electromagnetic fields $\vec E$ and $\vec B$ appearing as non-Abelian 
vector potentials. Pions couple to electromagnetism through a Goldstone-Wilczek
current that represents the baryon number of Skyrmions and gives rise to the 
decay $\pi^0 \rightarrow \gamma \gamma$. Similarly, magnons may couple to an 
analogue of the Goldstone-Wilczek current for baby-Skyrmions which induces a
magnon-two-photon vertex. The corresponding analogue of photon-axion 
conversion is photon-magnon conversion in an external magnetic field. The 
baryon number violating decay of Skyrmions can be catalyzed by a magnetic 
monopole via the Callan-Rubakov effect. Similarly, baby-Skyrmion decay can be 
catalyzed by a charged wire. For more than two flavors, the Wess-Zumino-Witten 
term enters the low-energy pion theory with a quantized prefactor $N_c$ --- the
number of quark colors. The magnon analogue of this prefactor is the anyon 
statistics angle $\theta$ which need not be quantized.

\end{abstract}
 
\maketitle
 
\newpage

\section{Introduction}

The concept of a spontaneously broken continuous global symmetry \cite{Nam60}
is important in many areas of physics. Irrespective of the details of the 
dynamics, Goldstone's theorem \cite{Gol61} predicts the existence of massless 
excitations just based upon the symmetry group $G$ and its unbroken subgroup 
$H$. At low energies the physics of the strong interaction is dominated by the 
lightest particles of QCD --- the pions \cite{Wei66}. In the chiral limit of 
zero quark masses, the pions are exactly massless Nambu-Goldstone bosons which 
stem from the $G = SU(2)_L \otimes SU(2)_R \otimes U(1)_B$ chiral symmetry of 
QCD that is spontaneously broken to $H = SU(2)_{L=R} \otimes U(1)_B$ at low 
temperatures. In the real world, on the other hand, the small non-zero quark 
masses give rise to a small pion mass and the pions are only pseudo-Goldstone 
bosons. The spontaneous breaking of the $G = SU(2)_s \cong SO(3)_s$ spin 
rotational symmetry group down to $H = U(1)_s \cong SO(2)_s$ plays an important
role in the condensed matter physics of ferro- and antiferromagnets. In 
particular, the low-energy physics of these systems is governed by associated 
massless Goldstone bosons --- the magnons or spin waves.

Since Goldstone bosons are massless, their dynamics dominate the low-energy
physics. These dynamics can be described by a low-energy effective theory 
involving only the Goldstone boson degrees of freedom. Irrespective of a 
concrete physical system, the Goldstone bosons are described by fields in the 
coset space $G/H$ \cite{Col69} whose dimension equals the number of
spontaneously broken generators. The low-energy effective theory must respect 
all symmetries of the underlying microscopic theory --- in particular those of
the global symmetry group $G$. Spontaneous symmetry breaking implies that the
Goldstone bosons interact weakly at low momenta. As a consequence, the 
Lagrangian of the effective theory can be constructed perturbatively as a 
derivative expansion, since terms with few derivatives dominate at low 
energies. A systematic effective field theory approach for describing Goldstone
boson physics --- chiral perturbation theory --- was developed for the pions in
QCD \cite{Gas85}, but is generally applicable to all Goldstone boson phenomena.
In condensed matter physics chiral perturbation theory has been applied to both
ferromagnetic \cite{Leu94,Hof99} and antiferromagnetic magnons 
\cite{Neu89,Fis89,Has90,Has93}.

In this paper we compare the low-energy physics of apparently quite different 
systems --- pions in QCD and magnons in ferro- and antiferromagnets. Since both
pions and magnons are Goldstone bosons, it is not surprising that they share a 
number of common features \cite{Bur98}. However, we find a correspondence 
between numerous pion and magnon phenomena that goes beyond what one might have
expected. In the standard model of particle physics the global $SU(2)_L$ 
symmetry of the pion effective theory turns into a local gauge symmetry once 
the weak interactions are taken into account. Similarly, the global $SU(2)_s$ 
spin symmetry of a magnon effective Lagrangian becomes local when 
electromagnetic interactions are included. This is quite surprising because 
electromagnetism, of course, results from gauging $U(1)_{em}$. Still, as 
pointed out by Fr\"ohlich and Studer \cite{Fro92}, for non-relativistic systems
the electromagnetic fields $\vec E$ and $\vec B$ appear as non-Abelian vector 
potentials of a local $SU(2)_s$ symmetry.

The topological Skyrmion excitations of the pion field \cite{Sky61} correspond 
to the baby-Skyrmions in ferro- and antiferromagnets \cite{Wil83}. When pions 
are coupled to the electromagnetic field, the Skyrme current, which describes 
baryon number, is generalized to the Goldstone-Wilczek current \cite{Gol81}. 
The Goldstone-Wilczek term in the low-energy effective Lagrangian gives rise to
the anomalous decay $\pi^0 \rightarrow \gamma \gamma$ of the neutral pion into 
two photons. When magnons are coupled to electromagnetism, the baby-Skyrmion 
current is replaced by an analogue of the Goldstone-Wilczek current. As we will
see, this term induces a magnon-two-photon vertex if the baby-Skyrmion carries 
electric charge. 

Baby-Skyrmions have been established experimentally in quantum Hall 
ferromagnets \cite{Tyc95}. They are well understood theoretically \cite{Gir98},
and, in particular, they do carry electric charge. Hence, magnon decay into two
photons should indeed occur in quantum Hall ferromagnets. Another particle that
can decay into two photons is the hypothetical axion \cite{Wei78,Wil78} --- the
Goldstone boson of a spontaneously broken Peccei-Quinn symmetry \cite{Pec77}. 
In order to make the axion visible in the laboratory, it has been proposed to 
use the axion-two-photon vertex in order to convert laser photons into axions 
inside a strong magnetic field \cite{Sik83,Bib87}. As we will see, there is a
condensed matter analogue of this effect --- namely photon-magnon conversion in
an external magnetic field.

The importance of baby-Skyrmions in the undoped antiferromagnetic precursors of
high-temperature superconductors \cite{Dzy88,Wie88,Shr90,Tim00} as well as 
topological mechanisms for superconductivity \cite{Wie92a} have also been 
discussed. Effective theories for holes doped into antiferromagnets were 
constructed in \cite{Wen89,Sha90}. It is possible that baby-Skyrmions in 
layered cuprate antiferromagnets carry hole quantum numbers. This hypothesis
can, at least in principle, be tested because it implies the decay of 
antiferromagnetic magnons into two photons. Again, photon-magnon conversion in 
an external magnetic field may make this process experimentally accessible. If 
baby-Skyrmions represent holes, the hedgehog structure of their staggered 
magnetization field may explain why antiferromagnetism is destroyed by doping 
\cite{Mar99}. If, in addition, baby-Skyrmions have an attractive interaction 
that can overcome their Coulomb repulsion, pairs of baby-Skyrmions may form 
already in the antiferromagnetic phase. It should be pointed out that the 
electromagnetic interactions of baby-Skyrmions are particularly interesting 
because the electromagnetic fields $\vec E$ and $\vec B$ appear as $SU(2)_s$ 
non-Abelian vector potentials. If baby-Skyrmion pairs form, it is conceivable 
that hole-doping of the antiferromagnet leads to their condensation and hence 
to superconductivity. It remains to be seen if this basic picture can be turned
into a more quantitative mechanism for high-temperature superconductivity. Our 
hope is that an effective field theory framework will be useful in this 
context.

There are further analogies between pions and magnons. The decay of both 
Skyrmions and baby-Skyrmions can be catalyzed by electromagnetic interactions. 
In particle physics this requires the presence of a magnetic monopole which can
catalyze baryon decay by the Callan-Rubakov effect \cite{Cal83,Rub81}. 
Similarly, baby-Skyrmion decay can be catalyzed by a charged wire sticking out
of a magnet. Furthermore, when several flavors $N_f \geq 3$ of quarks are 
considered, the Wess-Zumino-Witten term \cite{Wes71,Wit83} enters the 
low-energy pion effective theory. Interestingly, the prefactor of the 
Wess-Zumino-Witten term is quantized and corresponds to the number of quark 
colors $N_c$. In multi-layer quantum Hall ferromagnets the layer index plays 
the role of flavor \cite{Gir95}. An analogue of the Wess-Zumino-Witten term 
arises also in the corresponding generalized magnon effective theories. In this
case the corresponding prefactor is the anyon statistics angle $\theta$ for 
baby-Skyrmions which need not be quantized.

In this paper we concentrate on the general features of pion and magnon
effective theories with special emphasis on topological properties. In
particular, in the magnon context we do not limit ourselves to one specific
microscopic condensed matter system, but rather characterize the universal
properties of their low-energy description. Our framework is sufficiently 
general to incorporate condensed matter systems as different as single- or
multi-layer quantum Hall ferromagnets and antiferromagnetic precursors of
high-temperature superconductors. Many phenomena investigated here are well 
understood in condensed matter or particle physics, respectively. In fact, one 
purpose of this work is to underscore the numerous common features of 
apparently quite different condensed matter and particle physics systems and to
describe them in a universal effective field theory framework. Still, several
condensed matter phenomena discussed here --- namely magnon decay into two 
photons, photon-magnon conversion in a magnetic field, baby-Skyrmion decay 
catalyzed by a charged wire, or the effects caused by the Wess-Zumino-Witten
term for magnons --- have, as far as we know, not been discussed before.

The outline of this paper is as follows. In section 2 we compare the basic
low-energy effective theories of pions and magnons and we discuss their 
Skyrmion and baby-Skyrmion topological excitations. Section 3 contains a
discussion of the electromagnetic interactions of pions and magnons, in
particular, pion decay into two photons and photon-magnon conversion in an
external magnetic field. In section 4 the decay of Skyrmions and baby-Skyrmions
catalyzed by a magnetic monopole or a charged wire, respectively, is 
investigated. Section 5 generalizes the pion and magnon systems to several 
flavors and contains a discussion of the corresponding Wess-Zumino-Witten 
terms. Finally, section 6 contains a summary and our conclusions.

\section{Low-Energy Effective Theories for Pions and Magnons}

In this section we review the low-energy description of the dynamics of pions 
in $3+1$ dimensions as well as of ferro- and antiferromagnetic magnons in $2+1$
dimensions. We also discuss the topological solitons of these theories --- 
Skyrmions and baby-Skyrmions --- which can be quantized as bosons or fermions 
in $3+1$ dimensions, and as anyons in $2+1$ dimensions.

\subsection{Pions}

QCD with $N_c \ge 3$ colors and two massless quark flavors has a global chiral 
symmetry group $G = SU(2)_L \otimes SU(2)_R \otimes U(1)_B$ that is 
spontaneously broken down to the subgroup $H = SU(2)_{L=R} \otimes U(1)_B$ at 
low temperatures. Consequently, there are three massless Goldstone bosons --- 
the pions $\pi^+$, $\pi^0$, $\pi^-$ --- which are described by fields 
\begin{equation}
U(x) = \exp(2 i \pi^a(x) T^a/F_\pi)
\end{equation}
in the coset space $G/H = SU(2)_L \otimes SU(2)_R \otimes U(1)_B/SU(2)_{L=R} 
\otimes U(1)_B = SU(2)$. We have introduced the generators of $SU(2)$ such that
$\mbox{Tr}(T^a T^b) = \frac{1}{2} \delta_{ab}$. At low energies the pion 
dynamics are described by chiral perturbation theory. To lowest order, the 
corresponding Euclidean action is given by \cite{Gas85}
\begin{equation}
\label{action}
S[U] = \int d^4x \ \frac{F_\pi^2}{4} \mbox{Tr}[\p_\mu U^\dagger \p_\mu U].
\end{equation}
To leading order, $F_\pi$ is the pion decay constant. For simplicity, we have 
neglected an explicit symmetry breaking term due to non-zero quark masses. 
Indeed, the above action is invariant under global 
$G = SU(2)_L \otimes SU(2)_R$ transformations
\begin{equation}
U'(x) = L^\dagger U(x) R.
\end{equation}
The spontaneously selected constant vacuum field configuration $U(x) = \1$ is
invariant only under simultaneous transformations $L = R \in H = SU(2)_{L=R}$ 
on the left and on the right.

\subsection{Skyrmions}

Chiral perturbation theory is an expansion in small field fluctuations around 
the vacuum configuration $U(x) = \1$. While such perturbative fields can always
be continuously deformed into the vacuum configuration, general pion field 
configurations have non-trivial topological properties. In particular, the 
homotopy group $\Pi_3[SU(2)] = \Pi_3[S^3] = \Z$ implies that, at every instant 
in time, the pion field is characterized by an integer winding number
\begin{equation}
B = \frac{1}{24 \pi^2} \int d^3x \ \varepsilon_{ijk} \mbox{Tr}
\left[(U^\dagger \p_i U)(U^\dagger \p_j U)(U^\dagger \p_k U)\right].
\end{equation}
It was first realized by Skyrme that $B$ can be identified with the baryon 
number \cite{Sky61}. Hence, despite the fact that the pions themselves do not
carry baryon number, the topological solitons of the pion field --- the
Skyrmions --- can be regarded as baryons \cite{Wit83}. The baryon current
\begin{equation}
\label{bcurrent}
j_\mu = \frac{1}{24 \pi^2} \varepsilon_{\mu\nu\rho\sigma} \mbox{Tr}
\left[(U^\dagger \p_\nu U)(U^\dagger \p_\rho U)(U^\dagger \p_\sigma U)\right]
\end{equation}
is topologically conserved, i.e.\ $\p_\mu j_\mu = 0$, independent of the 
equations of motion. The dynamical properties of Skyrmions are not accessible 
in chiral perturbation theory because higher order terms in the action are then
equally important as the lowest-order term. Still, as we will see later, the 
fact that topologically non-trivial pion fields carry baryon number has 
profound consequences for the low-energy electromagnetic properties of pions.

In addition to $\Pi_3[SU(2)]$, the homotopy group $\Pi_4[SU(2)] = \Z(2)$ is
also non-trivial. Consequently, pion field configurations (now depending not 
only on space but also on time) fall in two distinct classes. Those that can be
continuously deformed into the vacuum configuration $U(x) = \1$ have a 
``winding'' number $\Sign[U] = 1$, while all other configurations have 
$\Sign[U] = - 1$. The topological number $\Sign[U]$ can be identified as the 
fermion permutation sign of the Skyrme soliton. For an odd number of colors 
(such as in the real world with $N_c = 3$) Skyrmions should be quantized as 
fermions because baryons then consist of an odd number of quarks. For even 
$N_c$, on the other hand, Skyrmions must be quantized as bosons. A pion field 
configuration $U(x)$ in which two Skyrmions interchange their positions as they
evolve in time has $\Sign[U] = - 1$ \cite{Wit83}. For odd $N_c$, i.e.\ for 
fermionic Skyrmions, the Pauli principle demands that such configurations 
contribute to the path integral with a negative sign. For even $N_c$, on the 
other hand, they should contribute with a positive sign. Hence, a factor 
$\Sign[U]^{N_c}$ appears in the pion path integral which takes the form
\begin{equation}
\label{pathintegral}
Z = \int {\cal D}U \exp(- S[U]) \ \Sign[U]^{N_c}.
\end{equation}
A configuration $U(x)$ in which a single Skyrmion rotates by $2 \pi$ during its
time evolution also has $\Sign[U] = - 1$. The inclusion of $\Sign[U]^{N_c}$ in 
the path integral is necessary to ensure that the Skyrmion has half-integer 
spin for odd $N_c$ and integer spin for even $N_c$.

\subsection{Antiferromagnetic Magnons}

Antiferromagnets are interesting condensed matter systems. In particular, the 
undoped precursors of high-temperature layered cuprate superconductors are
quantum antiferromagnets \cite{Sac03}. Microscopic models for these systems 
are, for example, the quantum Heisenberg model or the Hubbard model on a square
lattice at half-filling. Here we concentrate on models in $2+1$ dimensions 
which describe a single spatially 2-dimensional cuprate layer. In such systems 
at zero temperature the spin rotational symmetry group $G = SU(2)_s$ is 
spontaneously broken down to the unbroken subgroup $H = U(1)_s$ by the 
formation of a staggered magnetization. As a consequence, there are two 
massless Goldstone bosons --- the antiferromagnetic spin waves or magnons --- 
which are described by a unit-vector field 
\begin{equation}
\e(x) = (e_1(x),e_2(x),e_3(x)), \ \e(x)^2 = 1,
\end{equation}
in the coset space $G/H = SU(2)_s/U(1)_s = S^2$. Again, chiral perturbation 
theory describes the low-energy magnon dynamics by a Euclidean effective action
\cite{Cha89,Has91}, which to lowest order reads
\begin{equation}
S[\e] = \int d^2x \int_{S^1} dt \ \frac{\rho_s}{2} 
\left[\p_i \e \cdot \p_i \e + \frac{1}{c^2} \p_t \e \cdot \p_t \e\right].
\end{equation}
Here $\rho_s$ is the spin stiffness --- the analogue of $F_\pi^2$ in the pion 
case --- and $c$ is the spin wave velocity. The index $i \in \{1,2\}$ labels 
the two spatial directions, while the index $t$ refers to the time direction. 
We have compactified the Euclidean time dimension to a circle $S^1$ of 
circumference $\beta = 1/T$, which puts the system at a non-zero temperature 
$T$. Indeed, the magnon action is invariant under global rotations 
$O \in SO(3)_s \cong SU(2)_s = G$,
\begin{equation}
\e \ '(x) = O \e(x).
\end{equation}
The spontaneously selected constant vacuum field configuration $\e(x) = 
(0,0,1)$ is invariant only under transformations $O \in SO(2)_s \cong U(1)_s =
H$ in the unbroken subgroup. It is interesting that antiferromagnetic magnons 
have a ``relativistic'' energy-momentum dispersion relation
\begin{equation}
E = |\vec p| c,
\end{equation}
despite the fact that the underlying electron dynamics is non-relativistic. Of 
course, the spin wave velocity $c$ is smaller than the velocity of light.

\subsection{Ferromagnetic Magnons}

Ferromagnets are another class of interesting condensed matter systems which, 
for example, play a role in the context of the quantum Hall effect 
\cite{Gir98}. In particular, the Coulomb interaction in a spatially 
2-dimensional quantum Hall system favors a totally antisymmetric orbital wave 
function. Consequently, the spin wave function is totally symmetric and the 
system is ferromagnetically ordered. A microscopic model for ferromagnets is 
the quantum Heisenberg model. Just as in an antiferromagnet, in a ferromagnet 
the global spin rotational symmetry group $G = SU(2)_s$ is spontaneously broken
down to the subgroup $H = U(1)_s$, now by the formation of a uniform 
magnetization. Again, the corresponding Goldstone bosons --- in this case 
ferromagnetic magnons --- are described by a unit-vector field $\e(x)$ in the 
coset space $G/H = SU(2)_s/U(1)_s = S^2$. Unlike for antiferromagnets, the 
order parameter of a ferromagnet --- the uniform magnetization --- is a 
conserved quantity. This has interesting consequences for the low-energy 
physics of ferromagnetic magnons. In particular, their energy-momentum 
dispersion relation
\begin{equation}
E = \frac{\rho_s}{m}|\vec p|^2,
\end{equation}
is non-relativistic. Here $\rho_s$ is again the spin stiffness and $m$ is the
magnetization density. The lowest-order chiral perturbation theory Euclidean 
action for a ferromagnet is given by \cite{Leu94}
\begin{equation}
\label{faction}
S[\e] = \int d^2x \ \left[\int_{S^1} dt \ 
\frac{\rho_s}{2} \p_i \e \cdot \p_i \e -
i m \int_{H^2} dt \ d\tau \ \e \cdot (\p_t \e \times \p_\tau \e)\right].
\end{equation}
The second term on the right-hand side of this equation is of topological
nature. In order to write it in a manifestly $SU(2)_s$ invariant form, the
$(2+1)$-dimensional space-time has been extended into a fourth dimension with a
coordinate $\tau \in [0,1]$ which plays the role of a deformation parameter.
The manifold $H^2$ is a 2-dimensional hemisphere with the compactified 
Euclidean time $S^1$ as its boundary. The magnon field $\e(x)$ at physical 
space-time points $x \in \R^2 \times S^1$ is extended to a field $\e(x,\tau)$
in the 4-dimensional space-time $\R^2 \times H^2$ such that $\e(x,1) = \e(x)$ 
and $\e(x,0) = (0,0,1)$. Of course, the $(2+1)$-dimensional physics should be 
independent of how the field $\e(x,\tau)$ is deformed into the bulk of the 
fourth dimension. It should only depend on the boundary values $\e(x)$, i.e.\ 
on the magnon field in the physical part of space-time. This is possible 
because the integrand in the second term of eq.(\ref{faction}) is a total 
divergence closely related to the winding number of $\Pi_2[S^2] = \Z$. In 
fact, when the integration in eq.(\ref{faction}) over the hemisphere $H^2$ is 
replaced by an integration over a sphere $S^2$, the term
\begin{equation}
n = \frac{1}{4 \pi} \int_{S^2} dt \ d\tau \ 
\e \cdot (\p_t \e \times \p_\tau \e) \in \Pi_2[S^2] = \Z
\end{equation}
is an integer winding number. Hence, modulo an integer $n$, $S[\e]$ gets 
contributions only from the boundary of $\R^2 \times H^2$, i.e.\ from the 
$(2+1)$-dimensional physical space-time $\R^2 \times S^1$. Of course, one must 
still ensure that the integer contribution $n$ from the 4-dimensional bulk 
cancels. This is indeed the case, because the topological term $\frac{1}{4 \pi}
\int_{H^2} dt \ d\tau \ \e \cdot (\p_t \e \times \p_\tau \e)$ enters the action
with a prefactor 
\begin{equation}
\int d^2x \ 4 \pi i m = 4 \pi i M.
\end{equation}
Here $m$ is the magnetization density and
\begin{equation}
M = \int d^2x \ m
\end{equation}
is the total spin of the entire magnet and hence an integer or a half-integer. 
The 4-dimensional bulk ambiguity $4 \pi i M n$ in the action $S[\e]$ cancels in
the path integral. Due to the fact that $\exp(4 \pi i M n) = 1$, the factor 
$\exp(- S[\e])$ that enters the path integral is unambiguously defined. It is 
remarkable that consistency of the low-energy magnon path integral requires the
quantization of the total spin in integer or half-integer units.

\subsection{Baby-Skyrmions}

Just as pion fields support Skyrmions, both ferro- and antiferromagnetic
magnon fields support baby-Skyrmions. Baby-Skyrmions are solitons whose 
topological charge
\begin{equation}
B = \frac{1}{8 \pi} \int d^2x \ \varepsilon_{ij} 
\e \cdot (\p_i \e \times \p_j \e),
\end{equation}
again defined at every instant in time, is an element of the homotopy group 
$\Pi_2[S^2] = \Z$. In QCD the Skyrmion topological charge has been identified 
with baryon number. What physical quantity the baby-Skyrmion number $B$ 
represents depends on the specific ferro- or antiferromagnetic system in 
question. At this point we keep the discussion general and do not identify $B$ 
with a specific physical quantity. As before, the topological current
\begin{equation}
j_\mu = \frac{1}{8 \pi} \varepsilon_{\mu\nu\rho} 
\e \cdot (\p_\nu \e \times \p_\rho \e) 
\end{equation}
is conserved, i.e.\ $\p_\mu j_\mu = 0$, independent of the equations of motion.
As in the pion case, the detailed properties of baby-Skyrmions are not 
accessible in magnon chiral perturbation theory. Still, as we will discuss 
below, in cases where $B$ is related to the electric charge, the fact that 
baby-Skyrmions exist has profound consequences for the low-energy 
electromagnetic properties of magnons.

As in the pion case, there is another non-trivial homotopy group, $\Pi_3[S^2] =
\Z$, which is relevant for baby-Skyrmions. It implies that magnon fields 
(which now depend on both space and time) fall into distinct topological 
classes. The corresponding winding number is the Hopf number 
$H[\e] \in \Pi_3[S^2] = \Z$ which characterizes the braiding of baby-Skyrmion
paths in time. In $2+1$ dimensions particles can not only be quantized as 
bosons or fermions, but may have any spin and statistics. In particular, 
baby-Skyrmions can be quantized as anyons characterized by a statistics angle
$\theta$ \cite{Wil83}. The cases $\theta = 0$ and $\theta = \pi$ correspond to 
bosons and fermions, respectively. Including the Hopf term, the magnon path 
integral (both for ferro- and antiferromagnets) takes the form
\begin{equation}
\label{magnonPI}
Z = \int {\cal D}\e \ \exp(- S[\e]) \exp(i \theta H[\e]).
\end{equation}
The angle $\theta$ enters the magnon effective theory in a similar way as the
number of colors $N_c$ enters the pion effective theory. The value of $\theta$
must be determined for each individual underlying microscopic system. For 
example, for the antiferromagnetic quantum Heisenberg model it has been argued 
that no Hopf term is generated \cite{Wen88,Hal88,Dom88,Fra88,Rea89}, i.e.\
$\theta/2 \pi \in \Z$. Hence, in that case the baby-Skyrmions should be bosons.
It should be noted that Skyrmions in $3+1$ dimensions cannot be quantized as 
anyons. The homotopy group $\Pi_4[SU(2)] = \Z(2)$ allows only two cases --- 
bosons or fermions.

\section{Electromagnetism of Pions and Magnons}

In this section we couple the low-energy effective theories for pions and 
magnons to electromagnetism. In both cases, there are topological effects due
to Skyrmions or baby-Skyrmions. In the pion case these effects are described
by a Goldstone-Wilczek term which contains the vertex for the anomalous decay
$\pi^0 \rightarrow \gamma \gamma$ of a neutral pion into two photons. In cases
where the baby-Skyrmion topological charge $B$ is related to the electric 
charge there is an analogue of the Goldstone-Wilczek term for magnons which 
then gives rise to a magnon-two-photon vertex. This vertex can be used to 
convert photons into magnons in an external magnetic field.

\subsection{Pions and Photons}

At the quark level, the electric charge is given by
\begin{equation}
\label{charge}
Q = T^3_L + T^3_R + \frac{1}{2} B,
\end{equation}
where $T^3_L$ and $T^3_R$ are the diagonal generators of $SU(2)_L$ and 
$SU(2)_R$ and $B$ is the baryon number. Since $U(1)_B$ is not a subgroup of 
$SU(2)_L \otimes SU(2)_R$, it is not entirely straightforward to gauge the 
$U(1)_{em}$ symmetry of electromagnetism at the level of the pion effective 
theory. Naively, one would just replace ordinary derivatives with covariant 
ones. The electromagnetic covariant derivative of the pion field takes the form
\begin{equation}
D_\mu U(x) = \p_\mu U(x) + i e A_\mu [T^3,U(x)],
\end{equation}
where $A_\mu$ is the electromagnetic vector potential and $e$ is the electric
charge. The action is then given by
\begin{equation}
S[U,A_\mu] = \int d^4x \ \frac{F_\pi^2}{4} \mbox{Tr}[D_\mu U^\dagger D_\mu U].
\end{equation}
However, incorporating the covariant derivatives alone is not sufficient in 
order to gauge $U(1)_{em}$ correctly. Although the pions themselves do not 
carry baryon number, it is crucial to incorporate the baryon current in the 
effective theory since the quark charge $Q$ of eq.(\ref{charge}) contains the 
baryon number $B$. In particular, if one does not include the baryon current, 
the decay $\pi^0 \rightarrow \gamma \gamma$ does not happen in the effective 
theory. The baryon current of eq.(\ref{bcurrent}) is no longer conserved when 
ordinary derivatives are replaced with covariant ones, and the correct 
conserved baryon current is the Goldstone-Wilczek current 
\cite{Gol81,DHo84}
\begin{eqnarray}
\label{GWcurrent}
j^{GW}_\mu&=&\frac{1}{24 \pi^2} \varepsilon_{\mu\nu\rho\sigma} \mbox{Tr}
\left [(U^\dagger D_\nu U)(U^\dagger D_\rho U)(U^\dagger D_\sigma U)\right] 
\nonumber \\
&-&\frac{i e}{16 \pi^2} \varepsilon_{\mu\nu\rho\sigma} F_{\nu\rho} \mbox{Tr}
\left[T^3 (D_\sigma U U^\dagger + U^\dagger D_\sigma U)\right].
\end{eqnarray}
Since the quark charge $Q$ contains the baryon number $B$ with a prefactor
$1/2$, the Goldstone-Wilczek current should be coupled to the electromagnetic
field through an additional contribution to the action
\begin{equation}
\label{SGWem}
S_{GW}[U,A_\mu] = \frac{e}{2} \int d^4x \ A_\mu j_\mu^{GW}.
\end{equation}
The path integral of pions coupled to an external electromagnetic field then 
takes the form
\begin{equation}
Z[A_\mu] = \int {\cal D}U \exp(- S[U,A_\mu]) \ \mbox{Sign}[U]^{N_c} 
\exp(i S_{GW}[U,A_\mu]).
\end{equation}
One can now identify the vertex responsible for the decay 
$\pi^0 \rightarrow \gamma \gamma$. Putting 
$U(x) \approx 1 + 2 i \pi^0(x) T^3/F_\pi$, after partial integration the second
term in the Goldstone-Wilczek current of eq.(\ref{GWcurrent}) indeed yields the
vertex
\begin{equation}
{\cal L}_{\pi^0 \gamma \gamma}(x) = -i \frac{e^2}{32 \pi^2 F_\pi} \pi^0(x) 
\varepsilon_{\mu\nu\rho\sigma} F_{\mu\nu}(x) F_{\rho\sigma}(x),
\end{equation}
where
\begin{equation}
F_{\mu\nu}(x) = \p_\mu A_\nu(x) - \p_\nu A_\mu(x),
\end{equation}
is the electromagnetic field strength tensor. The above vertex is independent 
of the number of colors $N_c$. Indeed, in \cite{Bae01} it was shown that, in 
contrast to textbook knowledge, the $\pi^0 \rightarrow \gamma \gamma$ decay 
width does not depend on $N_c$ explicitly.

\subsection{Local $SU(2)_s$ Spin Symmetry of the Pauli Equation}

In order to make our paper self-contained, we include this subsection which is
based on work of Fr\"ohlich and Studer \cite{Fro92}. They realized that, up to 
order $1/M^3$ corrections (where $M$ is the electron mass), the 
non-relativistic Pauli equation (which results from reducing the Dirac equation
to its upper components) has a local $SU(2)_s$ spin symmetry. In the next
section, we will use this symmetry to construct an effective theory describing
the electromagnetic interactions of magnons and photons. The Pauli equation for
electrons interacting with the electromagnetic field, combined with the Pauli
equation for atomic nuclei, can be viewed as a condensed matter analogue of the
standard model of particle physics. Indeed, the electrodynamics of 
non-relativistic electrons and atomic nuclei interacting with photons, as 
complicated as it may be to solve, should, at least in principle, capture 
all phenomena in condensed matter. In practice it is impossible to derive 
emergent phenomena like the quantum Hall effect or high-temperature 
superconductivity from first principles of the underlying Pauli equation. 
Still, considering the underlying microscopic theory is useful, because its
symmetries are inherited by the low-energy effective theories that are used to
describe the various phenomena in question.

Up to corrections of order $1/M^3$ (and putting $\hbar = c = 1$), the Pauli 
equation describing the interaction of electrons with an external 
electromagnetic field $\Phi$, $\vec A$ can be cast into the form 
\cite{Fro92}
\begin{equation}
i (\p_t - i e \Phi + i \frac{e}{8 M^2} \vec \nabla \cdot \vec E
+ i \frac{e}{2M} \vec B \cdot \vec \sigma) \Psi = 
- \frac{1}{2 M}(\vec \nabla + i e \vec A - 
i \frac{e}{4 M} \vec E \times \vec \sigma)^2 \Psi.
\end{equation}
Here $\Psi$ is a 2-component Pauli spinor, and $\vec E = - \vec \nabla \Phi -
\p_t \vec A$ and $\vec B = \vec \nabla \times \vec A$ are the usual 
electromagnetic field strengths. The first two terms on the left-hand side form
the $U(1)_{em}$ covariant derivative familiar from electrodynamics. The third 
and fourth term on the left-hand side represent relativistic effects: the 
Darwin and Zeeman term, respectively. The first two terms on the right-hand
side again form an ordinary $U(1)_{em}$ covariant derivative, while the third 
term represents the relativistic spin-orbit coupling. Fr\"ohlich and Studer 
noticed a remarkable mathematical structure in the Pauli equation --- a local 
$SU(2)_s$ spin symmetry. Indeed, the above equation can be written as
\begin{equation}
i D_t \Psi = - \frac{1}{2 M} D_i D_i \Psi,
\end{equation}
with an $SU(2)_s \otimes U(1)_{em}$ covariant derivative given by
\begin{equation}
D_\mu = \p_\mu + i e A_\mu(x) + W_\mu(x).
\end{equation}
The components of the non-Abelian vector potential
\begin{equation}
W_\mu(x) = i W_\mu^a(x) T^a,
\end{equation}
(with $T^a = \frac{1}{2} \sigma^a$) can be identified as
\begin{equation}
\label{Wpot}
W_t^a(x) = \mu B^a(x), \ W_i^a(x) = \frac{\mu}{2} \varepsilon_{iab} E^b(x).
\end{equation}
Interestingly, the electromagnetic field strengths $\vec E$ and $\vec B$ enter 
the theory in the form of non-Abelian vector potentials $W_\mu$ of a local 
$SU(2)_s$ symmetry. The anomalous magnetic moment $\mu = g e/2 M$ of the 
electron (where, up to QED corrections, $g = 2$) plays the role of the 
non-Abelian gauge coupling. The Abelian vector potential $A_\mu$ is the one 
familiar from electrodynamics, except for a small contribution to the scalar 
potential due to the Darwin term, 
$A_t = - \Phi + (e/8 M^2) \vec \nabla \cdot \vec E$.

Fr\"ohlich and Studer's observation implies that in non-relativistic systems,
at least up to corrections of order $1/M^3$, spin plays the role of an 
internal quantum number analogous to flavor in particle physics. It should be 
pointed out that $SU(2)_s$ is not a local symmetry of the full microscopic 
theory underlying condensed matter physics. This follows because the energy 
density $\vec E^2 + \vec B^2$ of the electromagnetic field is invariant only 
under global (and not under local) $SU(2)_s$ transformations. In order to still
make use of the local symmetry, we separate the electromagnetic field into 
internal and external contributions. The internal contributions are responsible
for the complicated dynamics that turn electrons and atomic nuclei into ferro- 
or antiferromagnets. The internal fields have been integrated out and thus do 
not appear explicitly in the low-energy effective theory. External 
electromagnetic fields, on the other hand, are used to probe the physics of the
magnetic material and appear explicitly in the effective Lagrangian. Under 
these circumstances, the $\vec E^2 + \vec B^2$ contribution of the external 
field does not enter the dynamics, and the local $SU(2)_s$ symmetry is indeed 
realized. While it seems difficult to make these arguments quantitative, we 
think that they capture the essence of how magnets respond to external 
electromagnetic fields. Hence, although we expect the local $SU(2)_s$ symmetry 
to be only approximate in actual materials, in what follows we impose it as an 
exact symmetry. This allows us to identify the most important terms in the 
effective Lagrangian of magnons and photons. It should be noted that, despite
the intriguing mathematical structure, $SU(2)_s$ is not a gauge symmetry in the
usual sense. In particular, the non-Abelian vector potentials $W_\mu$ are 
nothing but the Abelian field strengths $\vec E$ and $\vec B$ and thus do not
represent independent physical degrees of freedom. Furthermore, there is no 
$SU(2)_s$ Gauss law. Consequently, ``gauge-variant'' states that carry a 
non-zero spin certainly still belong to the physical Hilbert space. Also, since
$SU(2)_s$ is not a true gauge symmetry, its spontaneous breakdown to $U(1)_s$ 
does not induce the Higgs mechanism. Since there are no $SU(2)_s$ gauge bosons
as independent degrees of freedom, the Goldstone boson mode cannot be 
incorporated as a longitudinal polarization state. Still, as we will see later,
some of the magnons pick up a mass due to their interactions with external 
electromagnetic fields.

\subsection{Magnons and Photons}

It is interesting to ask how magnons couple to photons. Despite the fact that
magnons are electrically neutral this question is non-trivial. The crucial
observation is that external electromagnetic fields couple to non-relativistic
condensed matter in the form of $SU(2)_s$ non-Abelian vector potentials. For 
antiferromagnetic magnons coupled to external $\vec E$ and $\vec B$ fields 
the effective action takes the form
\begin{equation}
\label{antif}
S[\e,W_\mu] = \int d^2x \int_{S^1} dt \ \frac{\rho_s}{2} 
\left[D_i \e \cdot D_i \e + \frac{1}{c^2} D_t \e \cdot D_t \e\right],
\end{equation}
with the covariant derivative 
\begin{equation}
D_\mu \e(x) = \p_\mu \e(x) + \e(x) \times \vec W_\mu(x).
\end{equation}
Similarly, for ferromagnetic magnons
\begin{eqnarray}
\label{ferro}
S[\e,W_\mu]&=&\int d^2x \ [\int_{S^1} dt \ \frac{\rho_s}{2} 
D_i \e \cdot D_i \e \nonumber \\
&-&i m \int_{H^2} dt \ d\tau \ \e \cdot (\p_t \e \times \p_\tau \e)
+ i m \int_{S^1} dt \ \e \cdot \vec W_t].
\end{eqnarray}
The last term is necessary to cancel the $SU(2)_s$ ``gauge'' variation of the 
second term. Both terms together are invariant.

Since magnons are electrically neutral, one might think that they do not couple
to the electromagnetic vector potential $A_\mu$. However, the case of pions in 
QCD has taught us that Goldstone bosons can couple indirectly to external 
fields through their topological excitations. For example, despite the fact 
that pions themselves do not carry baryon number, they couple to 
electromagnetism anomalously through the baryon number of their Skyrmion 
excitations. In fact, the decay $\pi^0 \rightarrow \gamma \gamma$ is entirely
due to this coupling. Similarly, magnons may couple to electromagnetism
indirectly through their baby-Skyrmion topological excitations. The magnon 
analogue of the Goldstone-Wilczek current is
\begin{equation}
\label{jGW}
j_\mu^{GW} = 
\frac{1}{8 \pi} \varepsilon_{\mu\nu\rho} \e \cdot (D_\nu \e \times D_\rho \e 
+ \vec W_{\nu\rho}),
\end{equation}
with the non-Abelian field strength given by
\begin{equation}
\vec W_{\mu\nu}(x) = \p_\mu \vec W_\nu(x) - \p_\nu \vec W_\mu(x) -
\vec W_\mu(x) \times \vec W_\nu(x).
\end{equation}
The condensed matter analogue of the standard model relation 
$Q = T^3_L + T^3_R + \frac{1}{2} B$ is
\begin{equation}
Q = - F,
\end{equation}
i.e.\ the electric charge in magnets is carried by the fermion number $F$ of 
the electrons. In this case there is no contribution from $T^3$ because both 
spin up and spin down electrons carry the same charge $- e$, while quarks of
flavor up and down have different electric charges. The fermion number of 
baby-Skyrmions is determined by the anyon angle as $F = B \theta/\pi = - Q$.
For example, if $\theta = \pi$, the baby-Skyrmions have electron quantum 
numbers, while for $\theta = 2 \pi$ they are bosons with charge $- 2e$. If the 
baby-Skyrmions carry an electric charge (i.e.\ if $\theta \neq 0$) the 
Goldstone-Wilczek current (times $- e \theta/\pi$) is the electric current, 
which hence couples to the electromagnetic field. Consequently, in analogy to 
QCD, a Goldstone-Wilczek term
\begin{equation}
S_{GW}[\e,A_\mu,W_\mu] = - \frac{e \theta}{\pi} \int d^3x \ A_\mu j_\mu^{GW},
\end{equation}
arises. The value of $\theta$ depends on the specific microscopic system in 
question.

Besides the Goldstone-Wilczek term, also the Hopf term $i \theta H[\e]$
contributes to the magnon action. The Hopf number $H[\e]$ is not invariant 
under local $SU(2)_s$ transformations. In fact, under a local transformation
$g(x) \in SU(2)_s$,
\begin{equation}
W_\mu'(x) = g(x)^\dagger(W_\mu(x) + \p_\mu) g(x),
\end{equation}
it changes by the winding number
\begin{equation}
n[g] = \frac{1}{24 \pi^2} \int d^3x \ \varepsilon_{\mu\nu\rho}
\mbox{Tr}\left[(g^\dagger \p_\mu g)(g^\dagger \p_\nu g)
(g^\dagger \p_\rho g)\right],
\end{equation}
and turns into
\begin{equation}
H[\e \ '] = H[\e] + n[g].
\end{equation}
This ``gauge'' variation indicates an anomaly in the baby-Skyrmion sector of 
the magnon effective theory, analogous to Witten's global anomaly \cite{Wit82}.
Since the local $SU(2)_s$ symmetry does not represent a true gauge symmetry of 
the underlying microscopic theory of electrons and atomic nuclei, this anomaly 
does not imply an inconsistency of the quantum theory, and thus need not 
necessarily be canceled. Still, the anomaly may be canceled by a Chern-Simons 
term
\begin{equation}
\label{ChernSimons}
S_{CS}[W_\mu] = \frac{1}{8 \pi^2} \int d^3x \ \varepsilon_{\mu\nu\rho}
\mbox{Tr}[W_\mu (\p_\nu W_\rho + \frac{2}{3} W_\nu W_\rho)].
\end{equation}
Like the Hopf term, the Chern-Simons term is not gauge invariant, and its 
gauge variation is given by
\begin{equation}
S_{CS}[W_\mu'] = S_{CS}[W_\mu] - n[g].
\end{equation}
Hence, $H[\e] + S_{CS}[W_\mu]$ is indeed invariant even against topologically
non-trivial gauge transformations. 

The magnon partition function takes the form
\begin{equation}
Z[A_\mu,W_\mu] = \int {\cal D}\e \ \exp(- S[\e,W_\mu]) 
\exp(i \theta H[\e]) \exp(i S_{GW}[\e,A_\mu,W_\mu]).
\end{equation}
Introducing small magnon fluctuations $m^a(x)$ ($a = 1,2$) around a (staggered)
magnetization in the 3-direction
\begin{equation}
\label{expansion}
\vec e(x) \approx (0,0,1) + \frac{1}{\sqrt{\rho_s}}(m^1(x),m^2(x),
- \frac{1}{2 \sqrt{\rho_s}}(m^1(x)^2 + m^2(x)^2)),
\end{equation}
one can identify the vertex responsible for the decay of a magnon into two 
photons
\begin{equation}
\label{vertex}
{\cal L}_{m \gamma \gamma}(x) = 
- i \frac{e \theta}{8 \pi^2 \sqrt{\rho_s}} m^a(x) 
\varepsilon_{\mu\nu\rho} F_{\mu \nu}(x) W^a_\rho(x).
\end{equation}
One photon is represented by the field strength tensor $F_{\mu \nu}$, while the
other one is contained in $W^a_\rho$. The experimental observation of magnon 
decay into two photons would unambiguously demonstrate that baby-Skyrmions 
indeed carry electric charge. Detecting this process inside a magnetic material
is certainly challenging, if not impossible. For example, for exactly massless 
magnons there is no phase space for the decay into photons. In the next 
subsection we discuss a set-up that may simplify the detection of the 
${\cal L}_{m \gamma \gamma}$ vertex.

In QCD the process $\pi^0 \rightarrow \gamma \gamma$ explicitly breaks the 
$G$-parity symmetry \cite{Lee56} through electromagnetic effects. At the level 
of the underlying microscopic standard model, the anomaly results from 
non-trivial transformation properties of the fermionic measure. In the 
low-energy effective theory, on the other hand, the measure is invariant under 
the symmetry, and the anomaly is represented by an explicit symmetry breaking 
term in the action.

It is interesting to ask what symmetry is anomalously broken by magnon decay
into two photons. The magnon analogue of $G$-parity is the $\Z(2)$ symmetry 
that turns $\e$ into $- \e$. Indeed, this symmetry is explicitly broken by the
magnon analogue of the Goldstone-Wilczek term. For antiferromagnetic magnons 
this electromagnetic effect is the only source of explicit $\Z(2)$ symmetry 
breaking. For ferromagnetic magnons, on the other hand, the topological term
$\vec e \cdot (\p_t \vec e \times \p_\tau \vec e)$ also breaks this symmetry.
The microscopic origin of this anomaly is the spin commutation relation
$[S_i,S_j] = i \varepsilon_{ijk} S_k$ which is satisfied for $\vec S$, but not
for $- \vec S$. It is interesting that, in contrast to the standard model, 
microscopically the anomaly does not originate from a non-trivial fermionic
measure, but from a non-trivial commutation relation. In both cases, the
breaking of the symmetry originates from quantum effects.

\subsection{Photon-Magnon Conversion in an External Magnetic Field}

We have seen that, just like pions, magnons can turn into two photons, 
provided that baby-Skyrmions carry electric charge. Hence, by studying 
magnon-photon interactions, one can learn something non-trivial about 
baby-Skyrmions. Observing 
magnon decay into two photons in a condensed matter experiment is a challenging
problem, and may even be impossible in practice. In this subsection we discuss 
a possible way of enhancing the magnon-two-photon process, which may make it 
more easily detectable. Again, the idea is inspired by particle physics --- 
namely by the conversion of photons into axions in an external magnetic 
field.\footnote{We thank W.~Bernreuther for reminding us of this process.} The 
axion \cite{Wei78,Wil78} is a hypothetical particle associated with the 
Peccei-Quinn mechanism \cite{Pec77} for solving the strong CP problem. Like 
pions and magnons, the axion is a Goldstone boson that can decay into two 
photons. Due to its very weak couplings, the axion is practically invisible and
has indeed not yet been found. However, in order to enhance axion visibility, 
interesting conversion experiments have been proposed \cite{Sik83,Bib87}. In 
particular, if one shines a very intense laser beam into a strong magnetic 
field, one can convert the laser photons into axions (provided that axions 
exist at all). This process makes use of the vertex for axion decay into two 
photons. One (real) photon is provided by the laser beam and the second 
(virtual) photon stems from the external magnetic field. Here we consider the 
magnon analogue of this process --- namely photon-magnon conversion in an 
external electromagnetic field. It remains to be seen if a laser beam shone 
into an antiferromagnetic precursor of a high-temperature superconductor or a 
quantum Hall ferromagnet will reveal the magnon-two-photon vertex 
experimentally. Here we provide some necessary theoretical background.

Let us first consider a ferromagnet in an external magnetic field $\vec B =
B \vec e_z$. Obviously, in order to minimize the energy, the vector $\vec e$ 
describing the uniform magnetization then aligns with the field $\vec B$. This
follows immediately from the term
\begin{equation}
m \ \vec e(x) \cdot \vec W_t(x) = m \mu \ \vec e(x) \cdot \vec B,
\end{equation}
in the effective magnon-photon Lagrangian of eq.(\ref{ferro}). Expanding the
magnon field as in eq.(\ref{expansion}), the term from above modifies the 
dispersion relation for ferromagnetic magnons \cite{Leu94} to
\begin{equation}
E = M_m + \frac{\rho_s}{m} |\vec p|^2,
\end{equation}
i.e.\ it leads to a magnon ``rest mass'' (or, more precisely, rest energy)
\begin{equation}
M_m = \mu B.
\end{equation}
Under these circumstances, the magnon-two-photon vertex of eq.(\ref{vertex})
can be written as
\begin{equation}
{\cal L}_{m \gamma \gamma}(x) = 
- i \frac{e \theta \mu B}{4 \pi^2 \sqrt{\rho_s}}
\left[m^1(x) B^1(x) + m^2(x) B^2(x)\right].
\end{equation}
Here $B^i$ ($i = 1,2$) represents the (real) laser photons, while the factor 
$B$ represents the (virtual) photons of the external magnetic field. It is 
important to ensure that the magnetic field component of the injected laser 
field is perpendicular to the direction of the external magnetic field.

Let us now consider an antiferromagnet in an external magnetic field. In order 
to minimize their energy, the spins in an antiferromagnet point antiparallel to
one another, but they should now also follow the external magnetic field. The 
best compromise to satisfy these competing requirements is achieved by a canted
state in which the staggered magnetization points perpendicular to 
the field. We now choose $\vec B = B \e_x$ and we again use the expansion of 
eq.(\ref{expansion}). For static fields, the contribution to the action that 
determines the canted state takes the form
\begin{equation}
D_t \e(x) \cdot D_t \e(x) = 
(\e(x) \times \vec W_t) \cdot (\e(x) \times \vec W_t) =
\mu^2 B^2 (1 - m^1(x)^2).
\end{equation}
Hence, in a magnetic field the magnon $m^1$ again picks up a mass 
$M_m = \mu B$, while the magnon $m^2$ remains massless. The dispersion relation
for the massive antiferromagnetic magnon is still ``relativistic'', i.e.\
\begin{equation}
E = \sqrt{M_m^2 + |\vec p|^2 c^2}.
\end{equation}
The magnon-two-photon vertex of eq.(\ref{vertex}) now takes the form
\begin{equation}
{\cal L}_{m \gamma \gamma}(x) = - i \frac{e \theta \mu B}
{4 \pi^2 \sqrt{\rho_s}} m^1(x) B^3(x).
\end{equation}
In this case, $B^3$ represents the laser photons. Again, the magnetic field 
component of the injected laser field should be perpendicular to the direction 
of the external magnetic field. Note that the laser photons can be converted
only into the massive magnon $m^1$.

\section{Skyrmion and Baby-Skyrmion Decay}

Skyrmions and baby-Skyrmions are topologically stable solitons. Still, when
they interact with external gauge fields, Skyrmions as well as baby-Skyrmions
can become unstable and decay. The decay of a Skyrme baryon in the pion 
effective theory can be induced by baryon number violating electroweak 
instantons through the 't Hooft anomaly. In addition, magnetic monopoles can
catalyze Skyrmion decay. There is no analogue of the 't Hooft anomaly for
baby-Skyrmions. However, baby-Skyrmion decay can still be catalyzed by the
condensed matter analogue of a magnetic monopole, an electrically charged wire.

\subsection{Pions, Skyrmions, and $W$-Bosons}

Since for magnons a local $SU(2)_s$ spin symmetry emerged somewhat 
unexpectedly, we now ask if there is an analogue of this for pions. Indeed, the
weak gauge interactions turn the global $SU(2)_L$ symmetry into a local one by 
coupling the pions to the non-Abelian $W$-boson field. In addition, the 
$U(1)_Y$ subgroup of $SU(2)_R$ is also gauged by coupling the pions to the 
Abelian $B$-bosons. The $SU(2)_L \otimes U(1)_Y$ symmetry then breaks 
spontaneously to the $U(1)_{em}$ symmetry of electromagnetism. The photon 
emerges as a linear combination of $W^3$ and $B$. In this subsection we 
concentrate on the $W$-bosons and thus we gauge only $SU(2)_L$ but not $U(1)_Y$
or $U(1)_{em}$. A more detailed discussion of the electroweak interactions of 
pions is contained in \cite{Bae01}.

Gauging $SU(2)_L$ is straightforward. One just replaces ordinary derivatives by
covariant derivatives
\begin{equation}
D_\mu U(x) = (\p_\mu + W_\mu(x)) U(x).
\end{equation}
Here $W_\mu = i g W_\mu^a T^a$ is the $SU(2)_L$ gauge field with gauge coupling
$g$ and field strength
\begin{equation}
W_{\mu\nu}(x) = \p_\mu W_\nu(x) - \p_\nu W_\mu(x) + [W_\mu(x),W_\nu(x)].
\end{equation}
The action now takes the form
\begin{equation}
\label{gaugedaction}
S[U,W_\mu] = \int d^4x \ \frac{F_\pi^2}{4} \mbox{Tr}[D_\mu U^\dagger D_\mu U],
\end{equation}
which is invariant under local transformations
\begin{equation}
U'(x) = L^\dagger(x) U(x), \ W_\mu'(x) = L^\dagger(x)(W_\mu(x) + \p_\mu) L(x).
\end{equation}
While the pion action $S[U,W_\mu]$ of eq.(\ref{gaugedaction}) is gauge 
invariant, the path integral as a whole is not. This is because
\begin{equation}
\mbox{Sign}[U'] = \mbox{Sign}[L U] = \mbox{Sign}[L] \ \mbox{Sign}[U].
\end{equation}
As pointed out by Witten \cite{Wit83} and by D'Hoker and Farhi \cite{DHo84}, 
the $SU(2)_L$ gauge variation of the fermion permutation sign of the Skyrmions
is a manifestation of Witten's global anomaly \cite{Wit82}. For odd $N_c$ the
gauged pion theory is inconsistent, unless the anomaly is canceled by 
additional fields. In the standard model the global anomaly is canceled by the
left-handed lepton doublet of neutrino and electron. For even $N_c$, on the 
other hand, the pure pion theory without leptons is anomaly-free and thus
consistent at the quantum level.

When $SU(2)_L$ is gauged, baryon number conservation is violated through the
't Hooft anomaly by electroweak instantons \cite{tHo76}. In this case, the 
Goldstone-Wilczek baryon number current \cite{Gol81,DHo84} takes the form
\begin{equation}
j^{GW}_\mu = \frac{1}{24 \pi^2} \varepsilon_{\mu\nu\rho\sigma} \mbox{Tr}
\left[(U^\dagger D_\nu U)(U^\dagger D_\rho U)(U^\dagger D_\sigma U)\right]
- \frac{1}{16 \pi^2} \varepsilon_{\mu\nu\rho\sigma} \mbox{Tr}
\left[W_{\nu\rho} (D_\sigma U U^\dagger)\right].
\end{equation}
Its divergence is given by
\begin{equation}
\p_\mu j^{GW}_\mu = - \frac{1}{32 \pi^2} \varepsilon_{\mu\nu\rho\sigma} 
\mbox{Tr}[W_{\mu\nu} W_{\rho\sigma}].
\end{equation}
Consequently, an electroweak gauge field with topological charge
\begin{equation}
Q = - \frac{1}{32 \pi^2} \int d^4x \ \varepsilon_{\mu\nu\rho\sigma} \mbox{Tr}
[W_{\mu\nu} W_{\rho\sigma}] \in \Pi_3[SU(2)_L] = \Z
\end{equation}
causes violation of baryon number conservation by $Q$ units. There is no 
analogue of the 't Hooft anomaly for magnons. The analogue of the 
Goldstone-Wilczek current for magnons is conserved independent of the form of 
the $SU(2)_s$ spin gauge field.

\subsection{Pions, Skyrmions, and Magnetic Monopoles}

The existence of magnetic monopoles was contemplated by Dirac as early as 1931
\cite{Dir31}. The standard model of particle physics does not contain 
magnetically charged particles and even Dirac did not believe in the existence
of magnetic monopoles at the end of his life \cite{Dir83}. Still, some 
extensions of the standard model --- for example, the $SU(5)$ grand unified 
theory --- contain very heavy 't Hooft-Polyakov monopoles which look like Dirac
monopoles from large distances. In the monopole core the $SU(5)$ symmetry is 
unbroken and quarks and leptons are indistinguishable there. As a consequence, 
baryons that enter the monopole core can reappear as leptons and thus the 
monopole itself can catalyze baryon decay. This is known as the Callan-Rubakov 
effect \cite{Cal83,Rub81}. In the $SU(5)$ grand unified theory, $B - L$ is 
conserved and thus baryon and lepton number are violated by the same amount. As
a result, $SU(5)$ monopoles also catalyze lepton decay.

The magnetic current of a monopole is given by
\begin{equation}
m_\sigma = \frac{1}{2} \varepsilon_{\mu\nu\rho\sigma} \p_\mu F_{\nu\rho},
\end{equation}
which measures the amount of violation of the Abelian Bianchi identity. In the 
presence of magnetic charge, the Goldstone-Wilczek current of 
eq.(\ref{GWcurrent}) is no longer conserved because
\begin{equation}
\label{dGW}
\p_\mu j_\mu^{GW} = - \frac{ie}{8 \pi^2} m_\sigma 
\mbox{Tr}\left[T^3(D_\sigma U U^\dagger + U^\dagger D_\sigma U)\right].
\end{equation}
For a magnetic monopole at rest at $\vec x = \vec 0$ we have
\begin{equation}
m_0(\vec x,t) = 4 \pi g \delta(\vec x), \ m_i(\vec x,t) = 0,
\end{equation}
where $g$ is the magnetic charge. In spherical coordinates 
$(r,\theta,\varphi)$, a vector potential describing this situation is given by
\begin{equation}
\vec A(\vec x) = g \frac{1 - \cos\theta}{r \sin\theta} \vec e_\varphi.
\end{equation}
This potential is singular along the negative $z$-axis, due to the Dirac 
string. Writing $U(x) = \exp(2 i \pi^0(x) T^3/F_\pi)$ and integrating 
eq.(\ref{dGW}) over space we obtain the rate of change of the baryon number as
\begin{equation}
\p_t B(t) = \frac{e g}{\pi F_\pi} \p_t \pi^0(\vec 0,t).
\end{equation}
Using the Dirac quantization condition $e g = 1/2$ one obtains
\begin{equation}
B(\infty) - B(- \infty) = \frac{1}{2 \pi F_\pi} 
\left[\pi^0(\vec 0,\infty) - \pi^0(\vec 0,- \infty)\right].
\end{equation}
Hence, if the neutral pion field $\pi^0(\vec 0)/F_\pi$ at the location of the
monopole rotates by $2 \pi n$, baryon number is violated by $n$ units.

\subsection{Magnons, Baby-Skyrmions, and Charged Wires}

The question arises if monopole catalyzed baryon decay has an analogue for 
mag\-nons. The corresponding analogue of the Goldstone-Wilczek current takes 
the form of eq.(\ref{jGW}). In analogy to the magnetic current we introduce
\begin{equation}
\vec m = \varepsilon_{\mu\nu\rho} D_\mu \vec W_{\nu \rho},
\end{equation}
which measures the amount of violation of the non-Abelian Bianchi identity. In
analogy to the QCD case, for non-vanishing $\vec m$ the Goldstone-Wilczek 
current is no longer conserved because
\begin{equation}
\p_\mu j_\mu^{GW} = \frac{1}{8 \pi} \vec m \cdot \vec e.
\end{equation}
As in the monopole case, we consider a point-like violation of the 
Bianchi-identity. The simplest example is
\begin{equation}
m^a(x) = 4 \pi g \ \delta^{a3} \ \delta(x).
\end{equation}
In this case, the $\delta$-function includes time, i.e.\ the violation of the
Bianchi-identity is event-like --- not particle-like. Hence, in the magnon 
theory the analogue of the magnetic monopole is a 3-dimensional instanton.
In complete analogy to the vector potential for a Dirac monopole one obtains
\begin{equation}
W^a_i(x) = g \ \delta^{a3} \ \frac{1 - \cos\theta}{r \sin\theta} e_{\varphi,i}.
\end{equation}
Introducing cylindrical space-time coordinates $\rho = r \sin\theta$, 
$\varphi$, and $t = r \cos\theta$ and using eq.(\ref{Wpot}) this equation
translates into
\begin{equation}
\vec E(\rho,t) = \frac{2 g}{\mu \rho} (1 - \frac{t}{\sqrt{t^2 + \rho^2}}) 
\vec e_\rho.
\end{equation}
In the far future the electric field vanishes, while in the distant past it 
takes the form $\vec E \sim 4 g \vec e_\rho/\mu \rho$. This is the electric 
field of a thin charged wire perpendicular to the 2-dimensional spatial plane 
with charge $8 \pi g/\mu$ per unit length. Hence, the instanton event describes
discharging a wire that leads out of the plane of the magnetic material. The 
discharging wire is the condensed matter analogue of the magnetic monopole. 
Similarly, a static charged wire (with time-independent charge) is the analogue
of the Dirac string. The resulting amount of baby-Skyrmion number violation is 
given by
\begin{equation}
B(\infty) - B(- \infty) = \frac{g}{2} e^3(0).
\end{equation}
It is clear that a wire sticking out of a magnet can transport electric charge 
out of the system. From the point of view of a 2-dimensional observer confined 
to the inside of the magnet this process violates charge conservation.

The requirement that the Dirac string emanating from a monopole is invisible 
implies the Dirac quantization condition. In particular, an Aharonov-Bohm 
scattering experiment on the Dirac string does not yield an observable 
interference pattern. In contrast to this, there is a non-trivial 
Aharonov-Casher effect, i.e.\ an observable interference pattern, when one 
scatters neutral particles with a non-zero magnetic moment off a static charged
wire \cite{Aha84,Fro92}. Using the concept of a local $SU(2)_s$ symmetry, this 
effect was also discussed by Anandan \cite{Ana89}.\footnote{We thank 
L.~Stodolsky for bringing this work to our attention.} There is no physical 
analogue of the Dirac quantization condition for charged wires. In particular, 
there is no reason why the mathematical analogue of the quantization condition 
should be realized in physical systems. After all, the amount of charge per 
unit length in the wire is under experimental control and need not be 
quantized.

\section{Generalization to Several Flavors}

There are interesting modifications of the low-energy effective theory for the
Goldstone bosons in QCD with more than two flavors. In particular, for 
$N_f \geq 3$ flavors the Wess-Zumino-Witten term arises with a quantized 
prefactor $N_c$. When one considers several coupled 2-dimensional layers of 
magnetic materials, the layer index may play the role of flavor. Then an 
analogue of the Wess-Zumino-Witten term may arise, however, its prefactor need 
no longer be quantized. 

\subsection{Pions, Kaons, and $\eta$-Mesons}

For QCD with $N_f \geq 3$ massless quarks the chiral symmetry group is 
$G = SU(N_f)_L \otimes SU(N_f)_R \otimes U(1)_B$ which is spontaneously broken 
down to the subgroup $H = SU(N_f)_{L=R} \otimes U(1)_B$. Hence, the Goldstone 
bosons are now described by fields in the coset space $G/H = SU(N_f)$. As a 
result, there are $N_f^2 - 1$ Goldstone bosons. For $N_f = 3$ there are 8 
Goldstone bosons: the 3 pions, 4 kaons, and the $\eta$-meson. The leading order
chiral perturbation theory action is still given by eq.(\ref{action}) as in the
$N_f = 2$ case. Since $\Pi_3[SU(N_f)] = \Z$ for any $N_f \geq 3$, the Skyrme 
and Goldstone-Wilczek currents of eqs.(\ref{bcurrent},\ref{GWcurrent}) also 
remain unchanged.

In contrast to the two flavor case, the homotopy group $\Pi_4[SU(N_f)]$ is 
trivial for $N_f \geq 3$. Hence, space-time-dependent Goldstone boson fields 
$U(x) \in SU(N_f)$ can then always be continuously deformed into the trivial 
field $U(x) = \1$. The question arises how the fermionic or bosonic nature of 
the Skyrmion manifests itself in the effective theory. Witten solved this
problem by introducing a fifth coordinate $x_5 \in [0,1]$ which plays the 
role of a deformation parameter \cite{Wit83}. He extended the 4-dimensional 
field $U(x)$ to a field $U(x,x_5)$ on a 5-dimensional hemisphere $H^5$ whose 
boundary $\p H^5 = S^4$ is (compactified) space-time, such that $U(x,0) = \1$ 
and $U(x,1) = U(x)$. One can now construct the Wess-Zumino-Witten term 
\cite{Wes71,Wit83} as
\begin{equation}
\label{WZW}
S_{WZW}[U] = \frac{1}{480 \pi^3 i} \int_{H^5} d^5x \ 
\varepsilon_{\mu\nu\rho\sigma\lambda} \mbox{Tr}\left[(U^\dagger \p_\mu U)
(U^\dagger \p_\nu U)(U^\dagger \p_\rho U)(U^\dagger \p_\sigma U)
(U^\dagger \p_\lambda U)\right].
\end{equation}
In analogy to the case of ferromagnetic magnons, the 4-dimensional Goldstone
boson physics should be independent of how the field $U(x,x_5)$ is extended to 
the bulk of the fifth dimension. It should only depend on the boundary values 
$U(x)$, i.e.\ on the Goldstone boson field in the physical part of space-time. 
Similar to the ferromagnetic magnon case, the integrand in eq.(\ref{WZW}) is a 
total divergence, and it is closely related to the winding number 
$\Pi_5[SU(N_f)] = \Z$. If the integration in eq.(\ref{WZW}) is performed
over a sphere $S^5$ instead of the hemisphere $H^5$, the result is the integer 
winding number of $U(x,x_5)$. Hence, modulo integers, $S_{WZW}[U]$ gets 
contributions only from the boundary of $H^5$, i.e.\ from the 4-dimensional 
physical space-time $S^4$. In order to ensure that the integer contribution 
from the 5-dimensional bulk cancels, $S_{WZW}[U]$ enters the path integral with
the quantized prefactor $N_c$, the number of colors,
\begin{equation}
\label{PI>2}
Z = \int {\cal D}U \exp(- S[U]) \exp(2 \pi i N_c S_{WZW}[U]).
\end{equation}
It should be noted that eq.(\ref{PI>2}) is the natural extension of 
eq.(\ref{pathintegral}) in the $N_f = 2$ case. Indeed, for $U(x) \in SU(2)$,
\begin{equation}
\exp(2 \pi i N_c S_{WZW}[U]) = \mbox{Sign}[U]^{N_c}.
\end{equation}
The argument of the Wess-Zumino-Witten term is a 5-dimensional Goldstone boson
field $U(x,x_5) \in SU(N_f)$ which reduces to a 4-dimensional $SU(2)$ field
$U(x)$ at the boundary of $H^5$. The argument of the sign factor, on the other
hand, is just the 4-dimensional field $U(x) \in SU(2)$. The Wess-Zumino-Witten 
term plays a similar role as $\mbox{Sign}[U]$ in the $N_f = 2$ case. In 
particular, for odd $N_c$ it ensures that the Skyrmion is quantized as a 
fermion with half-integer spin, while for even $N_c$ it is quantized as a boson
with integer spin \cite{Wit83}. 

\subsection{Antiferromagnetic Magnons with Several Flavors}

The question arises if the analogies between pions and magnons extend to 
several flavors. It is not clear, a priori, how to introduce additional flavors
of magnons. Instead of starting from concrete condensed matter systems, we let 
mathematics be our guide. We will ask later if the theories that arise in this 
way are realized in condensed matter physics. In QCD the two flavor case is 
generalized to several flavors by replacing the pion field $U(x) \in SU(2)$ by 
a Goldstone boson field $U(x) \in SU(N_f)$. What should replace the magnon 
unit-vector field $\e(x)$ in a generalization to several flavors? The goal is 
to generalize the $SU(2)_s$ spin rotational symmetry to $SU(N_f)$. Until now 
the magnon field $\e(x)$ lived in the coset space $S^2 = SU(2)/U(1) = CP(1)$. 
This suggests the generalization to $CP(N_f - 1)$ models. In particular, if a 
symmetry $G = SU(N_f)$ gets spontaneously broken to the subgroup 
$H = U(N_f - 1)$ the Goldstone bosons are described by fields in the coset 
space 
\begin{equation}
G/H = SU(N_f)/U(N_f - 1) = CP(N_f - 1).
\end{equation}
We will now consider low-energy effective theories describing such Goldstone 
bosons.

Goldstone bosons of $CP(N_f - 1)$ are described by $N_f \times N_f$ Hermitean 
projection matrices $P(x)$ that obey
\begin{equation}
P(x)^\dagger = P(x), \ \mbox{Tr} P(x) = 1, \ P(x)^2 = P(x).
\end{equation}
In the $N_f = 2$ (or $CP(1) = O(3)$) case the projection matrix is given by
\begin{equation}
\label{CP1O3}
P(x) = \frac{1}{2}(\1 + \e(x) \cdot \vec \sigma),
\end{equation}
where $\sigma^a = 2 T^a$ are the Pauli matrices. The lowest-order chiral
perturbation theory action for $CP(N_f - 1)$ antiferromagnetic magnons is 
given by
\begin{equation}
S[P] = \int d^2x \int_{S^1} dt \ \rho_s \left[\mbox{Tr}(\p_i P \p_i P) +
\frac{1}{c^2} \mbox{Tr}(\p_t P \p_t P)\right].
\end{equation}
This action is invariant under global special unitary transformations 
$g \in G = SU(N_f)$
\begin{equation}
P'(x) = g^\dagger P(x) g.
\end{equation}
The spontaneously selected vacuum field configuration $P(x) = 
\mbox{diag}(1,0,...,0)$ is invariant only under transformations $g$ in the 
unbroken subgroup $U(N_f - 1)$. Again, the magnons have a ``relativistic'' 
energy-momentum dispersion relation $E = |\vec p| c$.

\subsection{Ferromagnetic Magnons with Several Flavors}

Ferromagnetic magnons with several flavors arise in multi-layer quantum Hall
ferromagnets \cite{Gir95}. In these systems the layer index plays the role of 
flavor. Indeed, $CP(N_f - 1)$ effective theories have already been used to
describe these systems \cite{Eza97,Gho01}. For ferromagnetic magnons with 
several flavors the leading order chiral perturbation theory action is given by
\begin{equation}
\label{saction}
S[P] = \int d^2x \ \left[\int_{S^1} dt \ \rho_s \mbox{Tr}(\p_i P \p_i P) -
4 m \int_{H^2} dt \ d\tau \ \mbox{Tr}(P \p_t P \p_\tau P)\right].
\end{equation}
The second term on the right-hand side of this equation is again of topological
nature. The corresponding integrand is a total divergence closely related to 
the winding number of
\begin{equation}
\Pi_2[CP(N_f - 1)] = \Pi_2[SU(N_f)/U(N_f - 1)] = \Pi_1[U(N_f - 1)] = 
\Pi_1[U(1)] = \Z.
\end{equation}
Again, when the integration in eq.(\ref{saction}) over the hemisphere $H^2$ is 
replaced by an integration over a sphere $S^2$, the term
\begin{equation}
n = \frac{1}{\pi i} \int_{S^2} dt \ d\tau \ \mbox{Tr}(P \p_t P \p_\tau P)
\end{equation}
is an integer winding number. In order to ensure that the integer contribution 
$n$ from the 4-dimensional bulk cancels, the prefactor 
\begin{equation}
\int d^2x \ 4 \pi i m = 4 \pi i M
\end{equation}
must again be quantized, i.e.\ $M$ is an integer or a half-integer. 

\subsection{Baby-Skyrmions with Several Flavors}

Magnons with $CP(N_f - 1)$ low-energy dynamics also support baby-Skyrmions
because $\Pi_2[CP(N_f - 1)] = \Z$. The corresponding integer valued 
topological charge
\begin{equation}
B = \frac{1}{2 \pi i} \int d^2x \ \varepsilon_{ij} \mbox{Tr}(P \p_i P \p_j P)
\end{equation}
is constant in time because the topological current
\begin{equation}
j_\mu = \frac{1}{2 \pi i} \varepsilon_{\mu\nu\rho} \mbox{Tr}
(P \p_\nu P \p_\rho P)
\end{equation}
is conserved, i.e.\ $\p_\mu j_\mu = 0$.

In the QCD case we have seen that $\Pi_4[SU(2)] = \Z(2)$ while $\Pi_4[SU(N_f)]$
is trivial for $N_f \geq 3$, which gives rise to the Wess-Zumino-Witten term. 
In addition, $\Pi_5[SU(N_f)] = \Z$ leads to the quantization condition for the
prefactor $N_c$. Similarly, for magnons $\Pi_3[CP(1)] = \Z$ while 
$\Pi_3[CP(N_f - 1)]$ is trivial for $N_f \geq 3$. This gives rise to an
analogue of the Wess-Zumino-Witten term. However, since $\Pi_4[CP(N_f - 1)]$
is trivial, the prefactor of this term needs not to be quantized. This is
expected because the analogue of $N_c$ for magnons is the anyon statistics 
angle $\theta$ which is indeed not quantized.

Let us now construct the analogue of the Wess-Zumino-Witten term for magnons.
Since $\Pi_3[CP(N_f - 1)] = \{0\}$ for $N_f \geq 3$ space-time-dependent 
magnon fields $P(x) \in CP(N_f - 1)$ are topologically trivial and can always 
be continuously deformed into the constant field $P(x) = 
\mbox{diag}(1,0,...,0)$. As before, we introduce a fourth coordinate $\tau \in 
[0,1]$ which plays the role of a deformation parameter. First, we extend the 
3-dimensional field $P(x)$ to a field $P(x,\tau)$ on the 4-dimensional 
hemisphere $H^4$ whose boundary $\p H^4 = S^3$ is (compactified) space-time, 
such that $P(x,0) = \mbox{diag}(1,0,...,0)$ and $P(x,1) = P(x)$. The analogue 
of the Wess-Zumino-Witten term \cite{Jar85} takes the form
\begin{equation}
\label{magnonWZW}
S_{WZW}[P] = \frac{1}{4 \pi^2} \int_{H^4} d^4x \ \varepsilon_{\mu\nu\rho\sigma}
\mbox{Tr}(P \p_\mu P \p_\nu P \p_\rho P \p_\sigma P).
\end{equation}
Again, the 3-dimensional magnon physics should be independent of how the field 
$P(x,\tau)$ is deformed into the bulk of the fourth dimension. It should only 
depend on the boundary values $P(x)$, i.e.\ on the magnon field in the physical
part of space-time. In contrast to the QCD case where $\Pi_5[SU(N_f)] = \Z$, in
the magnon case $\Pi_4[CP(N_f - 1)]$ is trivial. Hence, if the integration in 
eq.(\ref{magnonWZW}) is performed over a sphere $S^4$ instead of the 
hemisphere $H^4$ the result simply vanishes. Thus, $S_{WZW}[P]$ gets 
contributions only from the boundary of $H^4$, i.e.\ from the 3-dimensional 
physical space-time $S^3$. In contrast to the QCD case, no bulk ambiguity 
arises and thus the prefactor of the Wess-Zumino-Witten term need not be
quantized. The path integral then takes the form
\begin{equation}
\label{magnonPI>2}
Z = \int {\cal D}P \exp(- S[P]) \exp(i \theta S_{WZW}[U]),
\end{equation}
where $\theta$ is again the (unquantized) anyon statistics angle. Indeed
eq.(\ref{magnonPI>2}) is the natural extension of eq.(\ref{magnonPI}) in the
$N_f = 2$ case. In particular, for $P(x)$ as in eq.(\ref{CP1O3}) we find
\begin{equation}
\exp(i \theta S_{WZW}[P]) = \exp(i \theta H[\e]).
\end{equation}
The argument of the Wess-Zumino-Witten term is a 4-dimensional magnon field 
$P(x,\tau) \in CP(N_f - 1)$ which reduces to a 3-dimensional $CP(1)$ field
$P(x)$ at the boundary of $H^4$. The argument of the Hopf term, on the other
hand, is just the 3-dimensional field $\e(x) \in S^2$. For $N_f \geq 3$ the 
Wess-Zumino-Witten term plays a similar role as the Hopf term for $N_f = 2$. In
particular, it determines that the baby-Skyrmion is quantized as an anyon with
statistics angle $\theta$.

It is interesting to consider the effects of gauging the $SU(N_f)$ symmetry.
Under an $SU(N_f)$ gauge transformation the magnon field transforms as
\begin{equation}
P'(x) = g^\dagger(x) P(x) g(x),
\end{equation}
and the corresponding non-Abelian gauge field transforms as
\begin{equation}
W_\mu'(x) = g^\dagger(x)(W_\mu(x) + \p_\mu) g(x).
\end{equation}
The gauged Goldstone-Wilczek current then takes the form
\begin{equation}
j_\mu^{GW} = 
\frac{1}{2 \pi i} \varepsilon_{\mu\nu\rho} \mbox{Tr}(P D_\nu P D_\rho P +
\frac{1}{2} P W_{\nu\rho}),
\end{equation}
which is again conserved. Hence, just as in the $N_f = 2$ case, there is no
analogue of the 't Hooft anomaly in the baryon number current. 

Let us also consider the modifications of the Wess-Zumino-Witten term when the 
$SU(N_f)$ symmetry is gauged. In the QCD context this has been done in 
\cite{Wit83,Cho84,Kaw84,Man85}. For magnons, the gauge variation of the 
Wess-Zumino-Witten term is given by
\begin{eqnarray}
&&S_{WZW}[P] - S_{WZW}[P'] = \frac{1}{4 \pi^2} \int d^3x \ 
\varepsilon_{\mu\nu\rho}[2 (\p_\mu g g^\dagger) \p_\nu P \p_\rho P P 
\nonumber \\
&&+ 2 (\p_\mu g g^\dagger) P (\p_\nu g g^\dagger) \p_\rho P P + \frac{2}{3} 
(\p_\mu g g^\dagger) P (\p_\nu g g^\dagger) P (\p_\rho g g^\dagger) P 
\nonumber \\
&&- (\p_\mu g g^\dagger) P (\p_\nu g g^\dagger) (\p_\rho g g^\dagger) P -
(\p_\mu g g^\dagger) \p_\nu P (\p_\rho g g^\dagger) P].
\end{eqnarray} 
Note that here the integration extends over the ordinary (compactified) 
$(2+1)$-dimensional space-time only. The gauge variation can be compensated by 
additional contributions to the Wess-Zumino-Witten term, which then takes the 
form
\begin{eqnarray}
S_{WZW}[P,W_\mu]&=&\frac{1}{4 \pi^2} \int_{H^4} d^4x \ 
\varepsilon_{\mu\nu\rho\sigma}
\mbox{Tr}(P \p_\mu P \p_\nu P \p_\rho P \p_\sigma P) \nonumber \\
&+&\frac{1}{4 \pi^2} \int d^3x \ \varepsilon_{\mu\nu\rho}
(P \p_\mu P W_\nu P W_\rho - \p_\mu P P W_\nu P W_\rho \nonumber \\
&+&2 P \p_\mu P \p_\nu P W_\rho + \frac{2}{3} P W_\mu P W_\nu P W_\rho +
P W_\mu P \p_\nu W_\rho).
\end{eqnarray}
In some multi-layer quantum Hall ferromagnets $SU(2)_s$ is a subgroup of 
$SU(N_f)$. In these cases, Fr\"ohlich and Studer's observations on the Pauli 
equation indeed imply that this subgroup should be turned into a local 
symmetry, with the electromagnetic field strengths $\vec E$ and $\vec B$ 
appearing as $SU(2)_s$ non-Abelian vector potentials. This is analogous to the 
standard model of particle physics where not the full chiral symmetry, but only
its $SU(2)_L \otimes U(1)_Y$ subgroup, is gauged. Besides the 
Goldstone-Wilc\-zek term, the gauged Wess-Zumino-Witten term contributes to 
anomalous electromagnetic effects. Again, this is analogous to QCD 
\cite{Bae01}.

\section{Summary and Conclusions}

In this paper we have compared the low-energy physics of pions and magnons and
we have found a surprising correspondence between various physical phenomena in
these apparently quite different particle and condensed matter physics systems.
For the two flavor case ($N_f = 2$) the analogies between pion and magnon 
physics are summarized in table 1. Similarly, table 2 summarizes the 
correspondences in the multi-flavor case $N_f \geq 3$. As we have seen, the 
topological structures in $(3+1)$-dimensional pion and $(2+1)$-dimensional 
magnon effective theories are very similar. Just based upon symmetry breaking 
patterns, low-energy effective field theory provides us with a universal 
framework for describing Goldstone boson physics, relating systems as different
as pions in QCD, undoped antiferromagnetic precursors of high-temperature 
superconductors, and single- or multi-layer quantum Hall ferromagnets.

\begin{table}
\begin{center}
\begin{tabular}{|c|c|c|}
\hline
Quantity & Pions & Magnons \\
\hline
\hline
global symmetry $G$ & $SU(2)_L \otimes SU(2)_R$ & $SU(2)_s$ \\
\hline
unbroken subgroup $H$ & $SU(2)_{L=R}$ & $U(1)_s$ \\
\hline
Goldstone field in $G/H$ & $U(x) \in SU(2)$ & $\e(x) \in S^2$ \\
\hline
coupling strength & pion decay constant $F_\pi$ & spin stiffness $\rho_s$ \\
\hline
propagation speed & velocity of light & spin wave velocity \\
\hline
topological solitons & Skyrmions & baby-Skyrmions \\
\hline
topological charge & baryon number $B$ & electron number $F$ \\
\hline
soliton homotopy & $\Pi_3[SU(2)] = \Z$ & $\Pi_2[S^2] = \Z$ \\
\hline
soliton statistics & bosons or fermions & anyons \\
\hline
statistics homotopy & $\Pi_4[SU(2)] = \Z(2)$ & $\Pi_3[S^2] = \Z$ \\
\hline
statistics factor & $\Sign[U]^{N_c}$ & $\exp(i \theta H[\e])$ \\
\hline
statistics parameter & number of colors $N_c$ & anyon angle $\theta/\pi$ \\
\hline
discrete symmetry & $G$-parity & $\e \rightarrow - \e$ symmetry \\
\hline
local symmetry & electroweak $SU(2)_L \otimes U(1)_Y$ & 
local $SU(2)_s \otimes U(1)_{em}$ \\
\hline
electromagnetic decay & $\pi^0 \rightarrow \gamma \gamma$ & 
magnon $\rightarrow \gamma \gamma$ \\
\hline
conversion in $\vec B$ field & photon-axion & photon-magnon \\
\hline
soliton decay catalyzer & magnetic monopole & charged wire \\
\hline
\end{tabular}
\end{center}
\caption{\it Analogies between pion and magnon physics.}
\end{table}

\begin{table}
\begin{center}
\begin{tabular}{|c|c|c|}
\hline
Quantity & Pions, Kaons, and $\eta$-Mesons & $N_f$ Magnon Flavors \\
\hline
\hline
global symmetry $G$ & $SU(N_f)_L \otimes SU(N_f)_R$ & $SU(N_f)$ \\
\hline
unbroken subgroup $H$ & $SU(N_f)_{L=R}$ & $U(N_f - 1)$ \\
\hline
Goldstone field in $G/H$ & $U(x) \in SU(N_f)$ & $P(x) \in CP(N_f - 1)$ \\
\hline
$SU(2)$ doublets & flavors $(u, d)$, $(c, s)$ & spins $(\uparrow, \downarrow)$ 
in layers \\
\hline
additional label & generation index & layer index \\
\hline
soliton homotopy & $\Pi_3[SU(N_f)] = \Z$ & $\Pi_2[CP(N_f - 1)] = \Z$ \\
\hline
soliton statistics & bosons or fermions & anyons \\
\hline
statistics homotopy & $\Pi_4[SU(N_f)] = \{0\}$ & $\Pi_3[CP(N_f - 1)] = \{0\}$ 
\\
\hline
WZW term & $\exp(2 \pi i N_c S_{WZW}[U])$ & 
$\exp(i \theta S_{WZW}[P])$ \\
\hline
WZW homotopy & $\Pi_5[SU(N_f)] = \Z$ & $\Pi_4[CP(N_f - 1)] = \{0\}$ \\
\hline
statistics parameter & quantized $N_c$ & unquantized $\theta$ \\
\hline
\end{tabular}
\end{center}
\caption{\it Analogies between pion, kaon, and $\eta$-meson physics and the
physics of magnons with several flavors.}
\end{table}

We have investigated low-energy effective theories for magnons without 
specifying a concrete magnetic material. The predictive power of the effective 
theory results from the fact that the details of the microscopic model enter 
the effective theory only through the values of some low-energy constants. For 
example, to leading order, the dynamics of magnons is determined by the spin 
stiffness $\rho_s$, as well as by the spin wave velocity $c$ (for 
antiferromagnets), or by the magnetization density $m$ (for ferromagnets). At
leading order, the electromagnetic properties of magnons are determined by the
local $SU(2)_s$ symmetry with the anomalous magnetic moment $\mu$ being 
the only low-energy parameter. Anomalous electromagnetic processes of magnons 
are due to the Goldstone-Wilczek term which, due to the relation $Q = - F$ 
between the electric charge and fermion number, is proportional to the anyon 
statistics parameter $\theta$. Determining the values of low-energy parameters 
(such as $\rho_s$, $c$, $m$, and $\theta$) for a concrete material is a 
non-trivial task. In general, one must use experiments in order to extract this
information. For simple model systems, such as the Heisenberg antiferromagnet, 
$\rho_s$ and $c$ have been obtained in very accurate quantum Monte Carlo 
calculations \cite{Wie92,Bea96}. 

In the context of the anomalous electromagnetism of magnons, the most 
interesting parameter is $\theta$, which determines the electric charge of 
baby-Skyrmions. For quantum Hall ferromagnets it is known that baby-Skyrmions 
have $\theta/\pi = \nu$, where $\nu$ is the Landau level filling fraction 
\cite{Gir98}. Hence, for $\nu = 1$ baby-Skyrmions are fermions and carry the 
charge of one electron. At fractional fillings, on the other hand, 
baby-Skyrmions, just like Laughlin quasi-particles, are anyons with fractional 
statistics and with fractional electric charge. 

For the antiferromagnetic precursor insulators of high-temperature 
superconductors the value of $\theta$ seems to be less clear. Let us discuss 
three different possible scenarios: 
\begin{itemize}
\item 
Since no Hopf term seems to be generated in the antiferromagnetic quantum 
Heisenberg model \cite{Wen88,Hal88,Dom88,Fra88,Rea89}, one might conclude that
$\theta = 0$. In that case baby-Skyrmions are neutral bosons and there is no
vertex for magnon decay into two photons. As a consequence, photon-magnon
conversion in a magnetic field is then impossible. Since neutral bosonic
baby-Skyrmions contain no net electrons, they are not directly affected by 
doping of the antiferromagnet. If $\theta$ indeed vanishes for the 
antiferromagnetic precursor insulators of high-temperature superconductors, one
cannot hope to learn anything about the destruction of antiferromagnetism by 
doping or about the preformation of Cooper pairs from the effective theory.
\item 
A more interesting scenario arises for $\theta = \pi$. Then the baby-Skyrmions 
are quasi-particles with electron quantum numbers. In that case, photon-magnon 
conversion in a magnetic field is possible. Doping forces a net number of 
baby-Skyrmions into the system, which may explain the destruction of 
antiferromagnetism due to the hedgehog form of the baby-Skyrmion's staggered
magnetization. Furthermore, investigating the forces between baby-Skyrmions 
mediated by magnon or photon exchange, one may hope to learn something about 
Cooper pair preformation within the effective theory. Still, in this case one
must understand the apparent conflict with the results of
\cite{Wen88,Hal88,Dom88,Fra88,Rea89}.
\item
The case $\theta = 2 \pi$ is also interesting. Then, in agreement with
\cite{Wen88,Hal88,Dom88,Fra88,Rea89}, no Hopf term is generated. However, the 
resulting bosonic baby-Skyrmions now have fermion number two and thus carry the
electric charge $- 2 e$. In that case, the baby-Skyrmions themselves represent 
preformed Cooper pairs. Again, doping forces baby-Skyrmions into the system, 
thus providing an explanation for the destruction of antiferromagnetism, and 
photon-magnon conversion is possible (even at a higher rate due to the larger 
value of $\theta$). Of course, if the baby-Skyrmions themselves are the 
preformed Cooper pairs, one cannot learn anything about the mechanism of their 
formation in the framework of the effective theory. Interestingly, this last 
scenario is analogous to QCD if one identifies electrons with quarks. In QCD
$N_c$ quarks are confined inside a baryon which manifests itself as a Skyrmion.
Although the prefactor $N_c$ of the Wess-Zumino-Witten term counts the number 
of baryon constituents, the effective theory does not shed any light on the 
mechanism by which quarks are confined inside baryons. Similarly, in the 
$\theta = 2 \pi$ scenario, two electrons are ``confined'' inside a preformed 
Cooper pair. The analogue of $N_c$ is $\theta/\pi = 2$ which again counts the 
number of constituents inside a baby-Skyrmion and appears as the prefactor of 
the Wess-Zumino-Witten term.
\end{itemize}

To decide if one of the above scenarios is realized in the antiferromagnetic 
precursors of high-temperature superconductors requires non-trivial insight
into these materials. On the one hand, one may hope to gain theoretical insight
from numerical simulations of the Hubbard model. On the other hand, one may
perform appropriate experiments. Motivated by the decay of the neutral pion 
into two photons, the analogy with magnons has led us to investigate if magnons
can decay into photons. Indeed, if baby-Skyrmions carry electric charge, the 
effective field theory predicts such a decay. The condensed matter analogue of 
photon-axion conversion, namely photon-magnon conversion in an external 
magnetic field relies on the magnon-two-photon vertex and may perhaps be 
realizable in experiments. If so, it should be observable in quantum Hall 
ferromagnets for which $\theta/\pi = \nu \neq 0$. If photon-magnon conversion 
was also observed in the antiferromagnetic precursors of high-temperature 
superconductors, this would show unambiguously that baby-Skyrmions in these
materials also carry electric charge. Even the value of $\theta$ can, at least 
in principle, be determined in this way. The practical feasibility of 
photon-magnon conversion experiments will be investigated in future studies.

Some condensed matter effects discussed in this paper were derived from 
analogies which are well understood in particle physics. One may also ask if we
can learn something new about particle physics, based on phenomena that are 
well understood in condensed matter. In this context, the layer index of a 
multi-layer quantum Hall ferromagnet suggests itself as a condensed matter
analogue of flavor, or (more precisely) of quark-lepton generation number. 
While the origin of flavor or generation number is a mystery in particle 
physics, the layer index in 2-dimensional quantum Hall systems obviously 
originates from the different locations of electrons in the third dimension. 
From this point of view, multi-layer quantum Hall systems seem to support the 
ideas behind brane worlds in which different flavors are localized in different
positions of an additional fourth spatial dimension \cite{Ark00}. In this 
context, attempts to use condensed matter language in order to describe 
fundamental physics should also be mentioned. This includes D-theory in which 
classical fields emerge dynamically by dimensional reduction of discrete 
variables \cite{Cha97}, emergent relativity \cite{Cha01}, effective gravity 
from quantum liquids \cite{Vol01}, as well as higher-dimensional analogs of the
quantum Hall effect \cite{Zha01,Fro00}, artificial light, and other recent 
developments \cite{Wen03}. Apparently, exploring the relations between particle
and condensed matter physics remains promising.

\section*{Acknowledgements}

We are grateful to W.~Bernreuther, S.~Chandrasekharan, J.~Fr\"ohlich, 
H.~Leutwyler, V.~Liu, M.~Troyer, and F.~Wilczek for very interesting 
discussions. O.~B.\ would like to thank the members of the theory group at Bern
University for their hospitality during two visits. This work is supported by 
funds provided by the Schweizerischer Nationalfond, by the U.\ S.\ Department 
of Energy (D.\ O.\ E.) under cooperative research agreement DF-FC02-94ER40818,
and by the European Community's Human Potential Program under contract 
HPRN-CT-2000-00145.

\end{document}